\documentstyle[twocolumn,prb,aps]{revtex}
\widetext

\begin{document} 

\twocolumn[\hsize\textwidth\columnwidth\hsize\csname @twocolumnfalse\endcsname
\draft
\preprint{}

\title{\sc Effects of magnetic field and disorder on electronic
properties of Carbon nanotubes}

\author{\sc Stephan Roche${\ }^{*}$ and Riichiro Saito${\ }^{\dagger}$}

\address{${\ }^{*}$ Department of Applied Physics, University of Tokyo, 7-3-1 Hongo,
Bunkyo-ku, Tokyo 113-8654, Japan.\\
${\ }^{\dagger}$ Department of Electronic Engineering, University of
Electro-communications, 1-5-1 Chofugaoka Chofu, Tokyo 182-8585, Japan
}
\date{\today}
\maketitle

\begin{abstract}
%\leftskip 54.8pt
%\rightskip 54.8pt
The electronic properties of carbon nanotubes are investigated in
presence of magnetic field perpendicular to the nanotube axis and
disorder introduced through site energy randomness. The magnetic field 
is shown to induce a metal-insulator transition for the metallic
(9,0),(12,0)... nanotubes in absence of disorder. It is also shown
that semiconducting nanotubes can become metallic with increasing
magnetic strength. Low disorder limit preserved the electronic
structure whereas metal-insulator transition remains even in the 
strong disorder limit. These results may find experimentally 
confirmation with tunneling spectroscopy measurements.
\end{abstract}


]
%\vspace{40pt}

%\vspace{20pt}

\section{Introduction}

%\vspace{20pt}

\hspace{\parindent}The discovery of carbon nanotubes (CNs)
 by Iijima \cite{CN-Ijima}, has sparked a tremendous amount
of activity and interest from basic research to applied
technology\cite{CN-basis}. These carbon-based structures
have spectacular electronic properties but the range of
their applications is much broader than electronics, and in
particular, due to potential marked mechanical properties,
they could become serious competitors to composite materials. 
CNs consists of coaxially rolled graphite sheets determined
by only two integers, and depending on the choice of $m$ and
$n$, a metallic, semiconducting or insulating behavior is
exhibited. Given their extremely small diameter (1 to 20
nanometer), CNs turn out to be serious candidates for the
next generation of microelectronic elementary devices often
referred to the unprecedented ``Nanotechnology Era''
\cite{Rohrer}.

%\vspace{20pt}

\hspace{\parindent}
The understanding of the pure CNs is now very
mature\cite{CN-basis}, and researches are now focusing on
the effects of topological (such as pentagon-heptagon pair
defects \cite{Charlier,CN-QD}) and chemical disorder on
``ideal properties'' in order to make CNs, operational devices
for microelectronics. Yet, transistor device has been
designed from CNs, and shown to operate at room temperature
\cite{CN-RTT}. Other mesoscopic effects, such as universal
conductance fluctuations(UCF), appearing at low temperature
have also been revealed in CNs\cite{langer96}. From Nanotubes,
1-dimensional (quantum wires) \cite{CN-QW} or zero
dimensional (quantum dots \cite{CN-QD}) molecular systems
have been conceived.

%\vspace{20pt}

\hspace{\parindent}To strengthen the usefulness of CNs, 
the understanding of the influence of defects in
nanostructures deserves a particular attention since at such
very small scale, any disruptions of local order may affect
dramatically the physical response of the nano-device. For
the implementation to efficient technological devices, one
needs seriously to understand, control and possibly confine
the magnitude of these inherent quantum fluctuations,
induced by disorder or magnetic field.

%\vspace{20pt}

\hspace{\parindent}Usually, classical theory enable 
an elegant description of electronic properties for
crystalline materials through the use of quantum mechanics
and symmetry principles. However, real materials always
enclosed impurities, dislocations, or more general form of
disorder which generally imply to go beyond usual paradigms
that have been used successfully for pure systems. To
investigate tunneling spectroscopy experiments or electronic
conductivity measurements, one may alternatively tackle the
problems of aperiodicity by using real space methods
\cite{ON}.

%\vspace{20pt}

\hspace{\parindent}Conductivity measurements have been performed on
bundles of single-walled carbon nanotubes by means of
scanning tunneling microscope (STM).\cite{tans98} By moving
the tip along the length of the nanotubes, sharp deviations
in the $I-V$ characteristics could be observed and related
to electronic properties predicted
theoretically. Notwithstanding, the scanning tunneling
experiment is a local measurement of current from the tip to
the nanotube surface and gives basically local informations
such as local density of states(LDoS)
\cite{CN-STM}. Further measurements of the electronic conductivity
involve the study of CN-based junctions. Finally, nanotubes have been 
recently shown to be nanometer-sized probes for imaging in chemistry and
biology, improving substantially the actual resolution of scanning probes\cite{CN-AFM}.

%\vspace{20pt}

\hspace{\parindent}Investigation of 
electronic properties are generally confined to a description of
the so-called electronic spectrum of a given material. From the
spectrum, one can typically distinguish a semiconductor from a
metallic sample by identification of a gap at Fermi energy. 
Within this framework, we will show that interesting effects are driven 
by the strength of the magnetic field, or energy fluctuations in the
site energies. In particular metal-insulator
transition will be studied, by considering the density of states,
when site randomness and magnetic filed act simultaneously.

%\vspace{20pt}

\section{Magnetic field induced metal-insulator transition}

%\vspace{20pt}

\hspace{\parindent}To evaluate the density of states (DoS) of the
carbon nanotubes, we use the recursion method \cite{Haydock}
which enables a study of structural imperfections and
disorder, as discussed by Charlier et al.\cite{Charlier} The
Green's function from which one estimates the DoS are
calculated as a continuous fraction expansion, which requires
to be properly terminated. Besides, finite imaginary part of
the Green's function is necessary to achieve numerical
convergence of the expansion if the system presents gaps in
the electronic structure. The DoS on a given initial state
$|\psi_{0}\rangle$ is evaluated from
$\langle\psi_{0}|G(z=E+i\eta)|\psi_{0}\rangle$ by tridiagonalization
of the corresponding tight-binding Hamiltonian. A proper
terminator has to be computed with respect to the asymptotic
behavior of recursion coefficients $a_{n},b_{n}$ associated
to the recursion procedure. For a metallic CN, the series
$a_{n},b_{n}$ exhibit well-defined limits $a_{\infty}$ and
$b_{\infty}$ enabling a simple termination of the Green's
function. For the semiconducting (10,0) CN, the series of
$b_{n}$ coefficients encompass small fluctuations confined
around 4.25 and 4.75 (in $\gamma$ units, with $\gamma$ the
coupling constant between neighboring carbon atoms). This
unveils the presence of a gap in the structure which may be
completely resolved by appropriate
terminator\cite{Turchi}. Furthermore, when dealing with
magnetic field which may lead to Landau levels structure,
suitable termination can be implemented \cite{Roche-PRB}.

%\vspace{20pt}

\hspace{\parindent}However, one notes that $\eta$ may 
also have some physical significance. Indeed, real materials
usually enclosed inherent disorder that lead to finite mean
free path, or scattering rates. Then, the first
approximation to account for such elastic/inelastic
relaxation times, is to introduced a finite imaginary part
of the Green's function (Fermi's Golden rule). Finally,
contrary to diagonalization methods for finite length
systems, a discussion about scaling properties is beyond the
scope of the recursion method. Finite size effects due to
boundary conditions will strongly affect physical
properties. For instance, close to a metal-insulator
transition, characteristic properties like localization
length will be driven by scaling exponents. Here, large
systems are used to achieve the convergence of continuous
fraction expansion before finite size effects occur. It is
equivalent to consider an infinite homogeneous
structure. Actually, local properties such as LDoS are
weakly affected by boundary effects when the nanotube is
sufficiently long.  In the following, we will keep a finite
imaginary part for the Green's function keeping in mind the
previous discussion.

%\vspace{20pt}

\hspace{\parindent}We consider a tight-binding description of the
graphite $\pi$ bands, with only first-neighbor C-C
interactions $\gamma=V_{pp\pi}$ which is taken as $-2.75$eV. 
The magnetic field is considered to be perpendicular to the
nanotube axis so that the potential vector is given by
$A=(0,\frac{LB}{2\pi}\sin
\frac{2\pi}{L}\cal{X})$\cite{Ajiki-1}
, with $a_{cc}=1.42\AA$, 
$L=\mid{\bf C}_{h}\mid =\sqrt{3}
a_{cc}\sqrt{n^2+m^2+nm}$, and the modulus of the chiral
vector defined by ${\bf C}_{h} =n{\bf a}_{1}+m{\bf
a}_{2}$\cite{CN-basis}. The effects of the magnetic field are driven by the
phase factors $\exp (\frac{ie}{\hbar c}\gamma_{ij})$
introduced in the hopping integrals between two sites ${\bf
R_{i}}=\{ {\cal X}_{i},{\cal Y}_{i}\}$ and ${\bf
R_{j}}$. They are generally defined by the potential vectors
through the Peierls substitution :

$$\gamma_{ij}=\frac{2\pi}{\varphi_{0}}\int_{R_i}^{R_j}{\bf
A}\cdot d{\bf r}$$

\noindent
where $\varphi_{0}=hc/e$ is the quantum flux. After simple algebra, 
assuming $\Delta {\cal X}= {\cal X}_{i}-{\cal
X}_{j}$ and $\Delta {\cal Y}= {\cal Y}_{i}-{\cal Y}_{j}$
one finds that the proper phase factors in our
case are given by\cite{CN-basis}
:

\begin{eqnarray}
\gamma_{ij}( \Delta {\cal X}\neq 0)&=&\bigl(\frac{L}{2\pi}\bigr)^{2}
B \frac{\Delta {\cal Y}}{\Delta {\cal X}}\nonumber\\
& & \times \biggl( -\cos \frac{2\pi}{L}({\cal X}+\Delta {\cal X})+
\cos(\frac{2\pi {\cal X}}{L})\biggr)\nonumber\\
\gamma_{ij}( \Delta {\cal X}= 0)&=&\bigl(\frac{L}{2\pi}\bigr)\Delta {\cal
Y}B\sin \frac{2\pi {\cal X}}{L}\nonumber\\
\end{eqnarray}

%\vspace{20pt}

%%%%%%%%%%%%%%%%Figure 1

\hspace{\parindent}On Fig. 1, we show the total density of
states (TDoS) at Fermi level as a function of the effective
magnetic field defined by $\nu=L/2\pi \ell$, where $\ell
=\sqrt{\hbar c/eB}$ is the magnetic length. For the two
metallic nanotubes (9,0) and (12,0), the TDoS at Fermi level
decreases as the magnetic strength is increasing. For higher
values of magnetic field, our results are in agreement which
was has been found by exact
diagonalizations\cite{CN-basis}.  As $\nu \to 1$, the TDoS, at Fermi
energy, approaches the same value ($\simeq 0.014$) as the
semiconducting (10,0) CN for zero magnetic field. 

%\vspace{20pt}

%%%%%%%%%%%%%%%%Figure 2

\hspace{\parindent}On Fig. 2, the
bold curve is the metallic CN (9,0), whereas the
dashed-bullet line stands for the semiconducting CN (10,0) both in
the absence of magnetic field B. The normal curve
corresponds to a metallic CN with magnetic field $\nu=0.8$. It 
can be seen that for $\nu=0.8$, the TDoS in the vicinity of the Fermi
level, presents a ``pseudo-gap'' (due to finite $\eta$) very similar to
the one of the (10,0) semiconducting nanotube for the same
value of $\eta=0.02$. Furthermore, we estimate the width
 $\Delta_{g}\simeq 1.1 eV$, which is in good agreement with typical
values found in experiments \cite{CN-STM}. This surprising result shows that the
magnetic field induces a metal-insulator transition. From $\nu=0$
to $\nu=0.8$, 
a continuous metal-insulator
transition can be drawn.

%%%%%%%%%%%%%%%%Figure 3

%\vspace{20pt}

\hspace{\parindent}The semiconducting case in Fig.~1 is also
interesting since a transition from semiconducting to
metallic is seen to occur for $\nu\sim 0.7$. On Fig.~3, the
effect of $\eta$ on the semiconducting case is
illustrated. As $\eta$ is reduced, the TDoS at Fermi level
(given in states/eV/graphite unit cell units) follows such
that
$\hbox{TDoS}(E_{F},\eta_{1})/\hbox{TDoS}(E_{F},\eta_{2})=\eta_{1}/\eta_{2}$.
Accordingly, a continuous transition, as $\eta$ tends to
zero, from insulator to metal (for $\nu$ of about 0.7) turns
out to be an inherent feature of the semiconducting
CN. Besides, Fig.~3 also shows the oscillation pattern that
is driven by the magnetic field.

%\vspace{20pt}

\hspace{\parindent}For the metallic CN, the normalized TDoS
 at Fermi level is given by $8/\sqrt{3}\pi a\gamma$ in the
unit of states/$\gamma$/(length along the
nanotube)\cite{CN-basis}, with $a=2.46\AA$, so that the
expected real TDoS$(E_{F})$ for (9,0) nanotube is given by
0.08168 states/$\gamma$/(graphite unit cell). From our
results, we find that TDoS$(E_{F},\eta=0.02)=0.08238$ and
TDoS$(E_{F},\eta=0.01)=0.0812$ which are qualitatively in
good agreement with the theoretical result. We have checked
that the integral density of states (IDoS) is normalized.

%%%%%%%%%%%%%%%%Figure 4

%\vspace{20pt}

\hspace{\parindent}For the (9,0) CN,  Ajiki
and Ando \cite{Ajiki-1} predicted that the DoS, 
for magnetic field parallel to the nanotube
axis and in the framework of $k \cdot p$ approximation, should exhibit
$\varphi_{0}$-periodic oscillations ($
\varphi_{0}=ch/e$ the quantum flux). In our
case, for $\nu=0$ to $\nu=0.8$, the finite density
of states in the vicinity of Fermi level (Fig. 4) 
is deepened and finally reach the same value
as the semiconducting CN for same value of $\eta$. Afterwards,
TDoS undergoes as a function of $\nu$ non-periodic oscillations, as it
can be seen when $\nu$ increases from 1 to 1.4.

%\vspace{20pt}

\hspace{\parindent}To conclude this first part, from our results it
seems that magnetic field leads to a oscillatory behavior of the TDoS at Fermi
level between metallic and semiconducting electronic states. This
effect has been illustrated symmetrically on semiconducting and metallic
nanotubes for broad range of magnetic field strengths.

%\vspace{20pt}

\section{Combination of randomness and magnetic field effects}

%\vspace{20pt}

\hspace{\parindent}As mentioned previously, it is of great necessity
to analyze the stability of physical patterns that emerge from perfect 
nanotubes, when introducing disorder. So far much consideration
has been paid to the so-called heptagon-pentagon pair defects since
their presence is to be expected in any junction devices \cite{Charlier}. Hereafter,
we will rather focus on the effect of Anderson-type disorder on
spectral properties, investigating to which extent, randomness is able 
to modify the previously unveiled pattern. We concentrate our study in the
metallic CN case, restricting our discussion to (9,0) CN, given that
similar patterns were observed in others metallic ones.

%\vspace{20pt}

\hspace{\parindent}The effect of randomness is now considered on the
site energies of the tight-binding Hamiltonian with 
$\sum_{n_{x},n_{y}}\varepsilon_{n}|n_{x},n_{y}\rangle
\langle n_{x},n_{y}|$. The site energies are
 randomly chosen in the interval $[-W/2,W/2]$, with uniform probability
distribution. Accordingly, the strength of disorder is measured by $W$. To evaluate the
TDoS, we use an average on typically 100 different
configurations.

%\vspace{20pt}

\hspace{\parindent}The TDoS as a function of Fermi energy
for $\nu=0.8$ with different values of randomness, in the
vicinity of the Fermi energy ($E_{F}=0$ eV) has been studied.  Values of
randomness strength W=0, 0.5, 1,2 and 3 have been considered
and are given in $\gamma$ units. The electronic structure at Fermi level is
affected in several ways. First, it is interesting to point
out that disorder up to W=1.0 does not break localization
properties of nanotubes. Unlike usual 1D system where the slightest
disorder induces Anderson localization mechanism, CN may be considered 
as marginal 1D quantum nanostructures where quantum confinement takes
place but existing phenomena in 1D-structures undergo specific
alterations. The effect of disorder being one of them.
One has to reach higher values of
disorder strengths ($W\sim 1.5$) to break suck localization.

%%%%%%%%%%%%%%%%Figure 5

%\vspace{20pt}

\hspace{\parindent}If we
consider the enlargement of the DoS over the entire
bandwidth (Fig. 5), one clearly sees that from W=1.5, all the
one-dimensional van-Hove singularities have
disappeared. This is a signature that quantum confinement
has been disrupted by disorder. Besides, for W=3 bandwidth
is enlarged by 1/4 of the total bandwidth, which is a rather
small increase of the bandwidth. Probably the localization
length is then smaller than the circumference of the CN.

%\vspace{20pt}

\hspace{\parindent}We then consider the evolution of DoS as a function
of $\nu$ on  Fig. 6 for $W=0,0.5,1.5$. The case $W=0$
 is given and compared with low-disorder limit $W=0.5$. For each value 
of $\nu$, we plot a different LDoS on a given site. The effect of low
disorder is shown not to affect significantly the general
metal-insulator pattern discussed earlier. This is also the case for
stronger disorder. However for disorder
width as large as 1.5 (as shown on Fig. 6), LDoS in the low-magnetic field regime
is strongly affected by the appearance of
strong fluctuations between different LDoS, whose period
further increases with magnetic strength. As $\nu$
approaches 1, one recovers a basic pattern that magnetic
field will induced metal-insulator transition. Averaging on
some 10 LDoS reduces the amplitude of such fluctuations, but 
 the TDoS at Fermi energy for low disorder limit is increased
(bold-curve on Fig.6).

%%%%%%%%%%%%%%%%Figure 6

%\vspace{20pt}

\hspace{\parindent}We believe that this effect of fluctuations may have
significant consequences on the electronic properties of CN,
for instance affecting the ideal properties of nanoscale
metal-semiconductor contact devices \cite{CN-QW} made from
CN. This should also be considered carefully in relation to
UCF. Indeed, it is generally assumed that UCF reflects the
microscopic random potential where electrons are
propagating. The pattern of the fluctuations of conductance
as a function of Fermi energy or magnetic field, being quite
random but reproducible and fluctuating from sample to
sample. The only common feature is that the fluctuations are
of order of $e^{2}/h$ independently of sample quality, size,
and so forth.

%\vspace{20pt}

\hspace{\parindent}To some extent, LDoS from one site to another is
also related to the local potential around a given carbon atom, and if 
general UCF may be seen as a consequence of fluctuations in the
potential distribution, universal fingerprints may also emerge in local spectral
properties. Besides, as the LDoS can be directly related to tunneling
current from tips to the surface, such fluctuations 
may be indirectly related to STM experiments. Calculation of Kubo
conductivity by means of real-space recursion method \cite{ON} may
also lead to valuable information about transport properties \cite{SR-RS}.

%\vspace{20pt}

\section{Conclusion}

%\vspace{20pt}

\hspace{\parindent}We have shown several patterns occurring in the
electronic spectra of metallic or semiconducting carbon nanotubes and
induced by magnetic field and disorder. Magnetic field was shown to
lead to a continuous metal-insulator transition in both kind of CN, whereas
disorder was shown not to modify qualitatively the abovementioned
pattern. Strong fluctuations of LDoS as a function of site environment 
and magnetic field were found and may be of importance when designing
junction devices.

\acknowledgments
SR is indebted to the European Commission and the
Japanese Society for Promotion of Science (JSPS) for joint
financial support (Contract ERIC17CT960010), and to Prof. T. 
Fujiwara from Department of Applied Physics of Tokyo
University for his kind hospitality.  Part of the work by RS
is supported by a Grant-in Aid for Scientific Research
(No.~10137216) from the Ministry of Education and Science of
Japan.

\vfill\eject

\section*{Figures captions}

%\vspace{20pt}

%\noindent
%Fig. 1. Recursion coefficients $b_{n}$ for the Carbone Nanotube (10,0) 
%presenting a gap at Fermi level.

%\vspace{20pt}

\noindent
Fig. 1. Evolution of the TDoS for (9,0) and (10,0) CNs as a function of
magnetic field.

%\vspace{20pt}

\noindent
Fig. 2. Comparison of TDoS for a metallic (9,0) CN (bold-line), a metallic
(9,0) CN under the influence of magnetic field. $\nu=0.8$ (bullet-line) and
a semiconducting (10,0) CN (solid-line) for the same value of $\eta$.

%\vspace{20pt}

\noindent
Fig. 3. Effect of the finite imaginary part of Green's function ($\eta$) on
TDoS for a semiconducting (10,0) CN.

%\vspace{20pt}

\noindent
Fig. 4. TDoS for metallic (9,0) CN for different values 
of $\nu$.

%\vspace{20pt}

%\noindent
%Fig. 6.  TDoS as a function of energy for different values of disorder
%and a magnetic field corresponding to $\nu=0.8$, in the vicinity of
%Fermi level

%\vspace{20pt}

\noindent

Fig. 5. TDoS of (9,0) CN as a function of energy for
different values of disorder and a magnetic field
corresponding to $\nu=0.8$

%\vspace{20pt}

\noindent
Fig. 6. TDoS and LDoS of (9,0) CN as a function of magnetic
field for disorder strengths W=0, 0.5, and 1.5.

%\vspace{20pt}

%\noindent
%Fig. 6. Solid line gives the TDoS in the W=0 case, whereas bullet-line 
%represent the LDoS for different sites as a function of magnetic
%strength, and dashed line is an average on 10 LDoS.

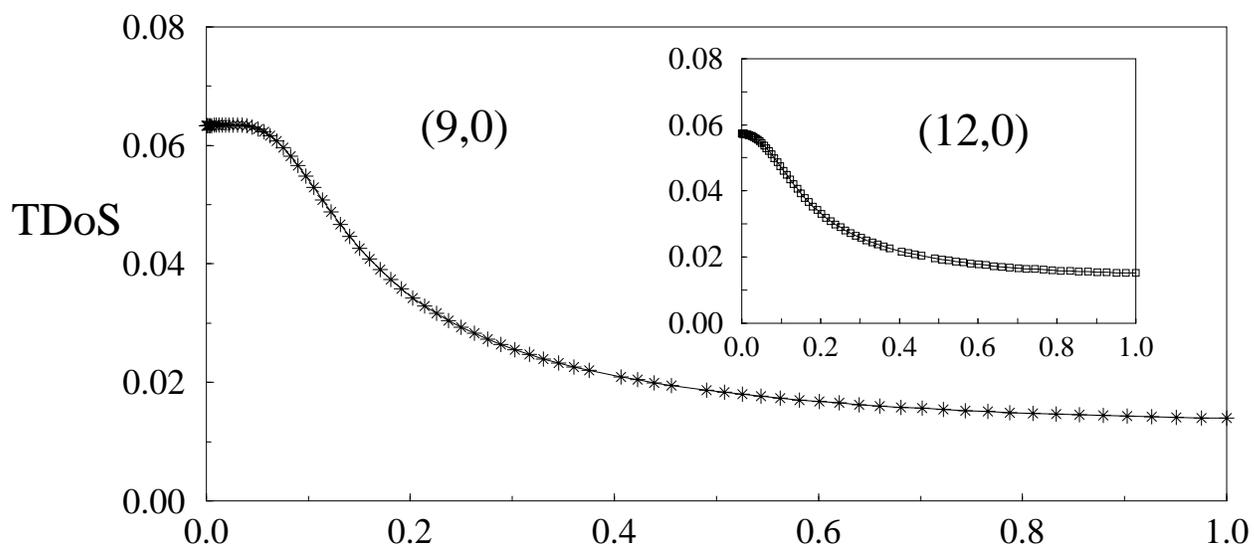
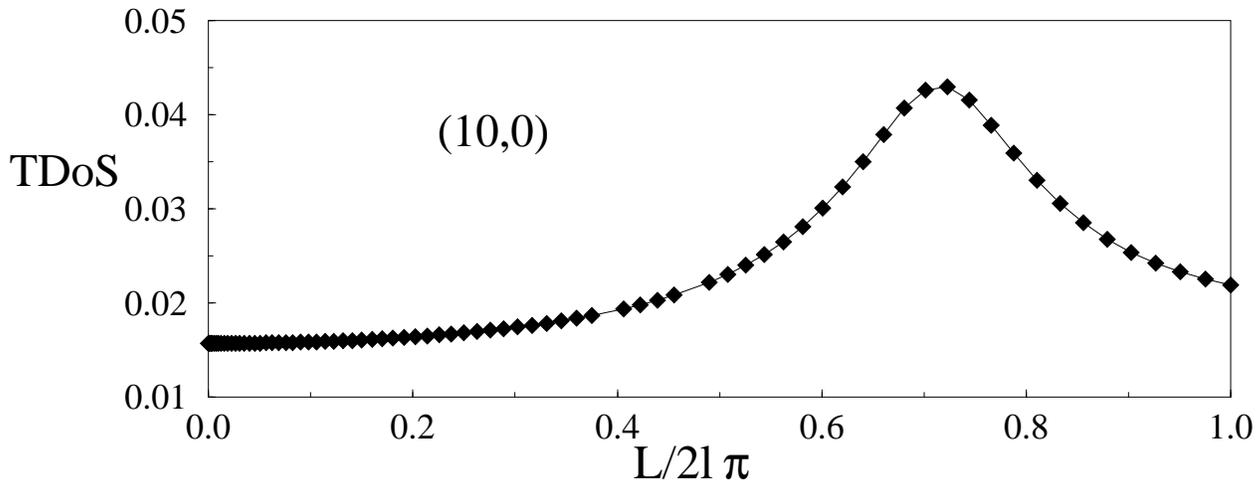
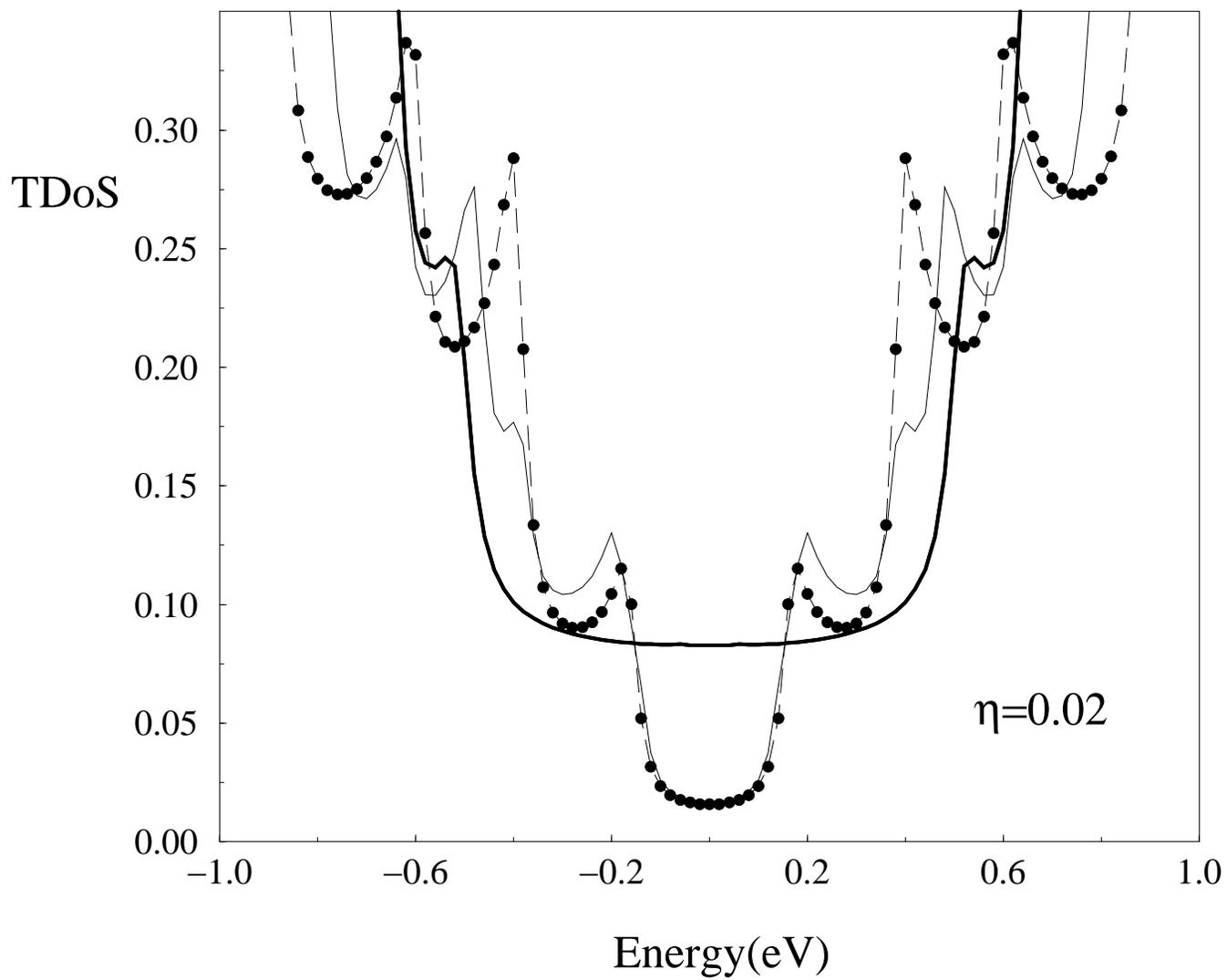
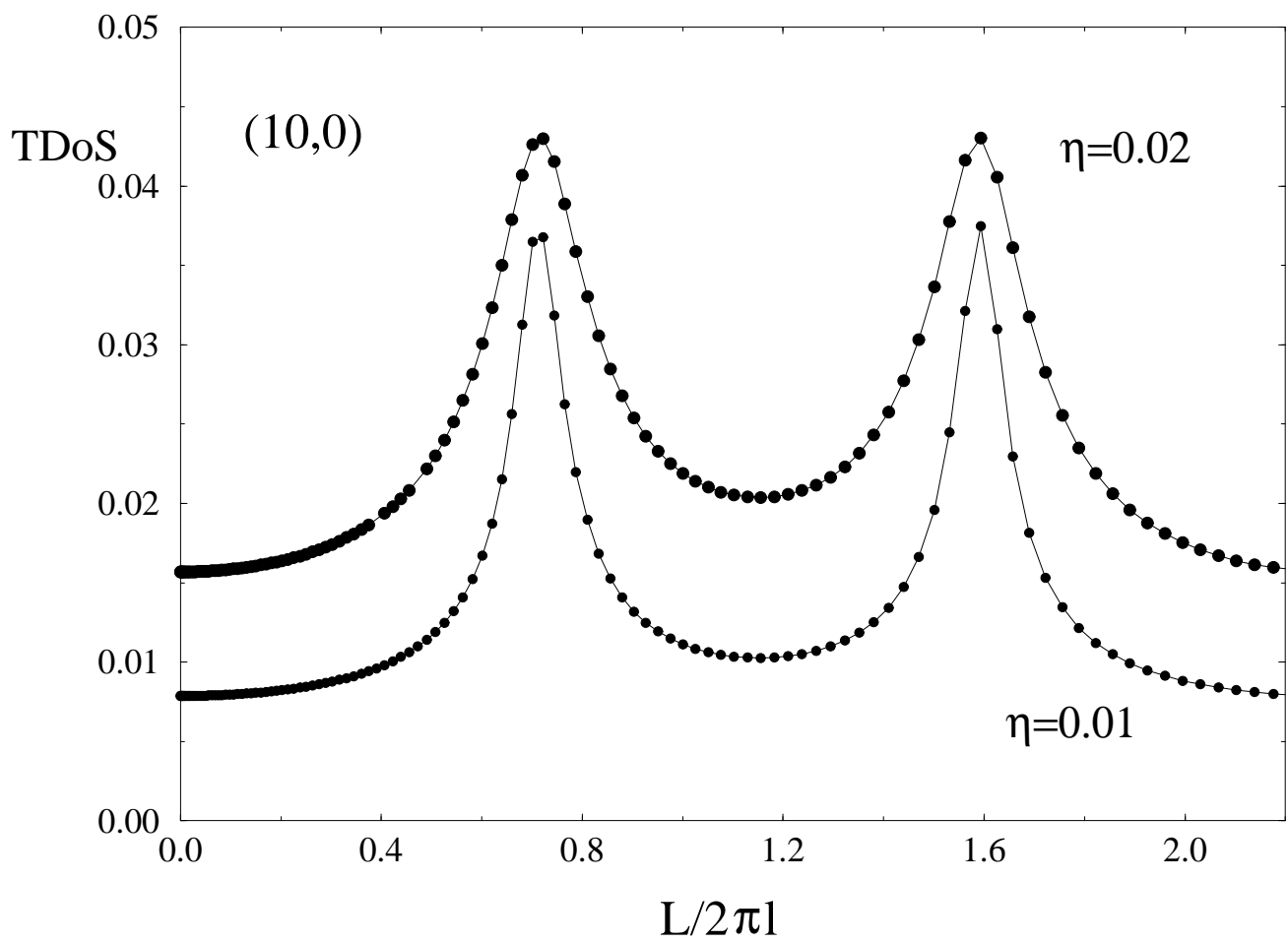
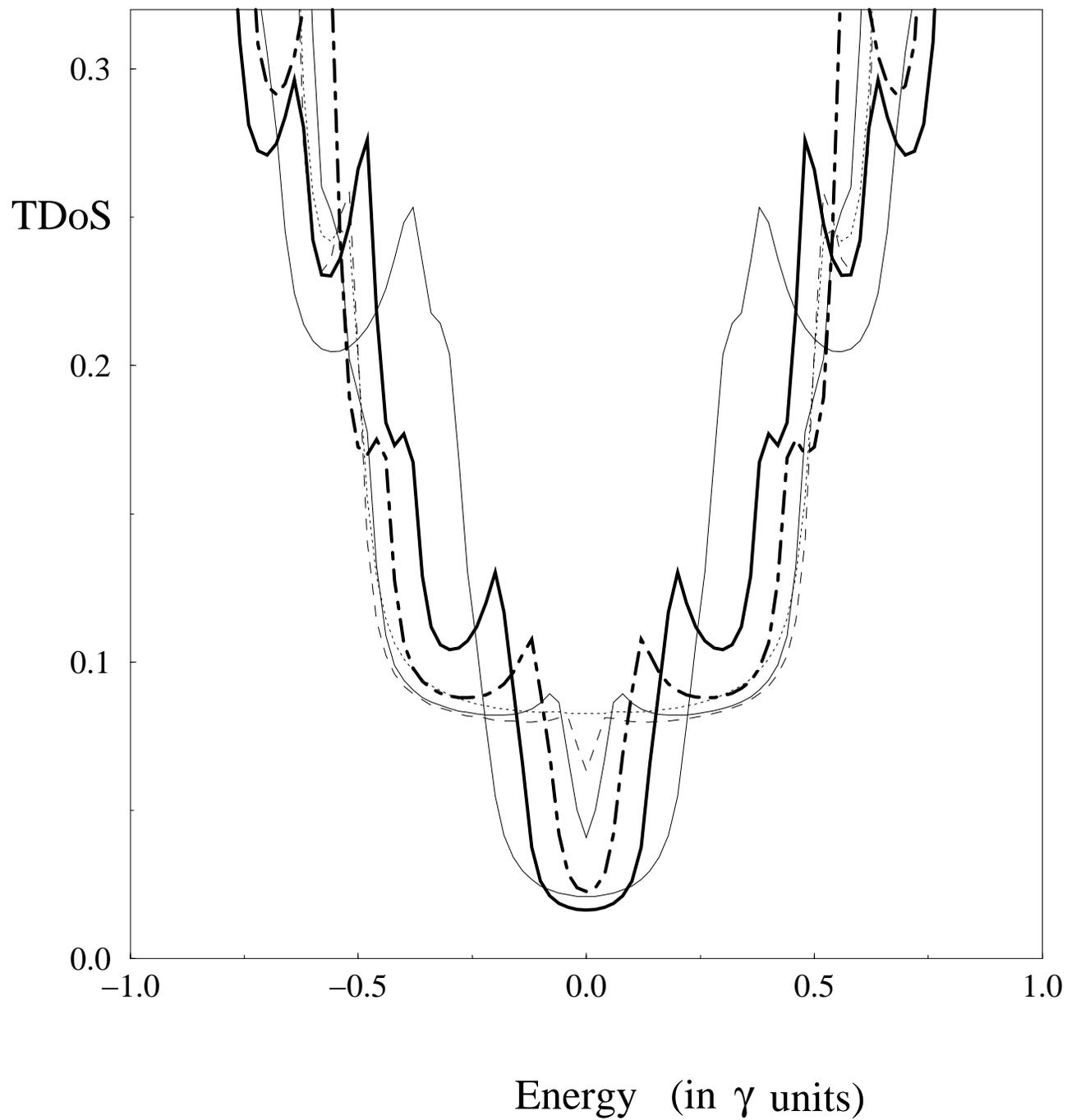
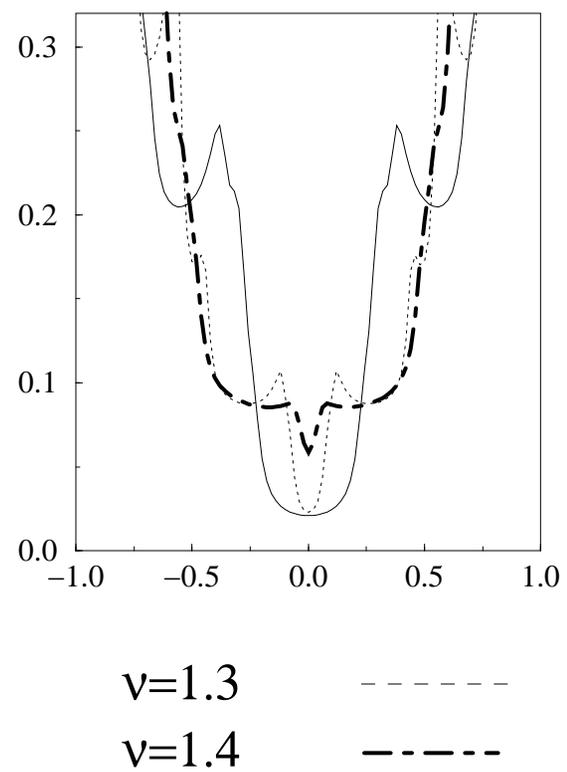
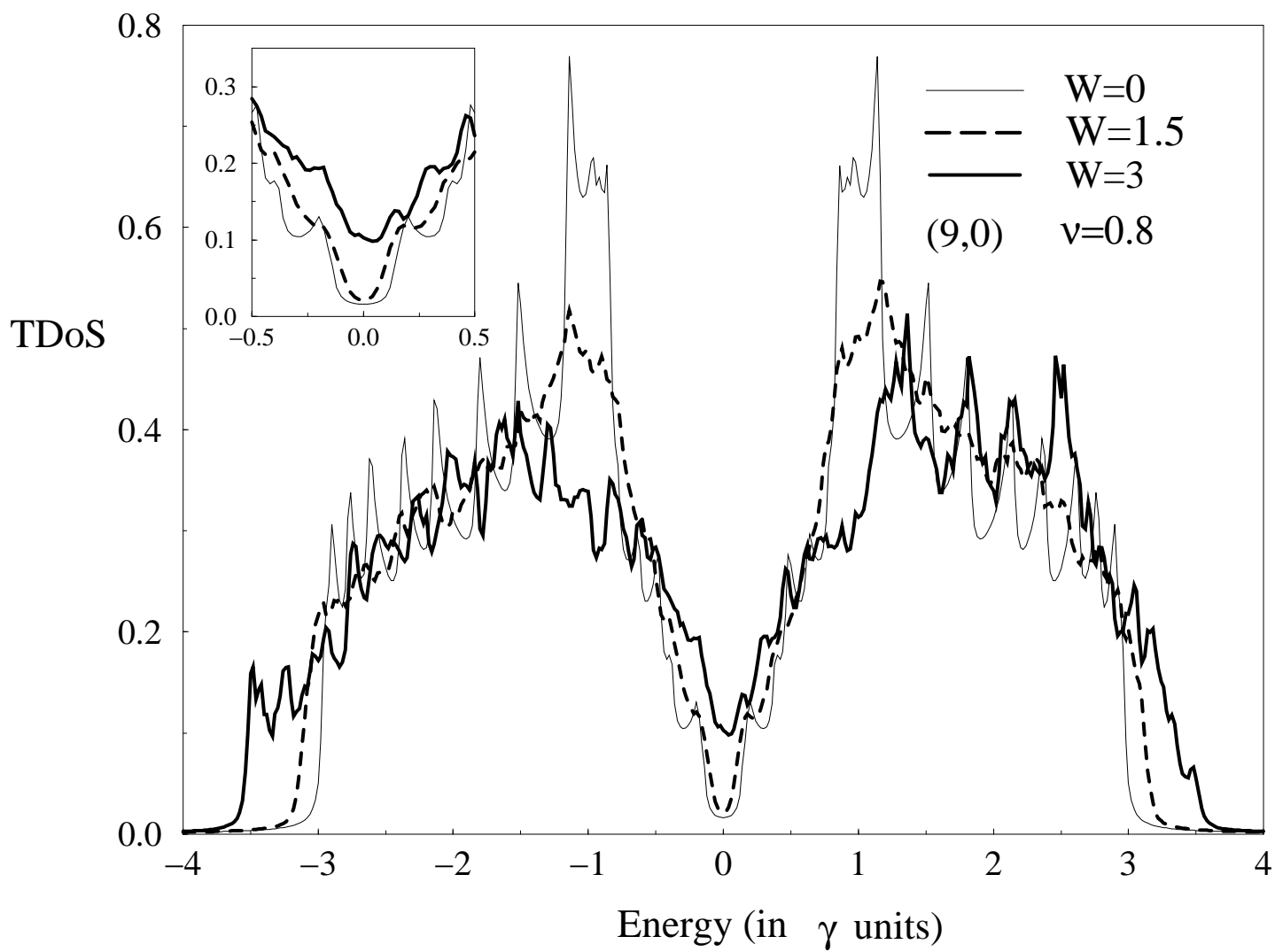
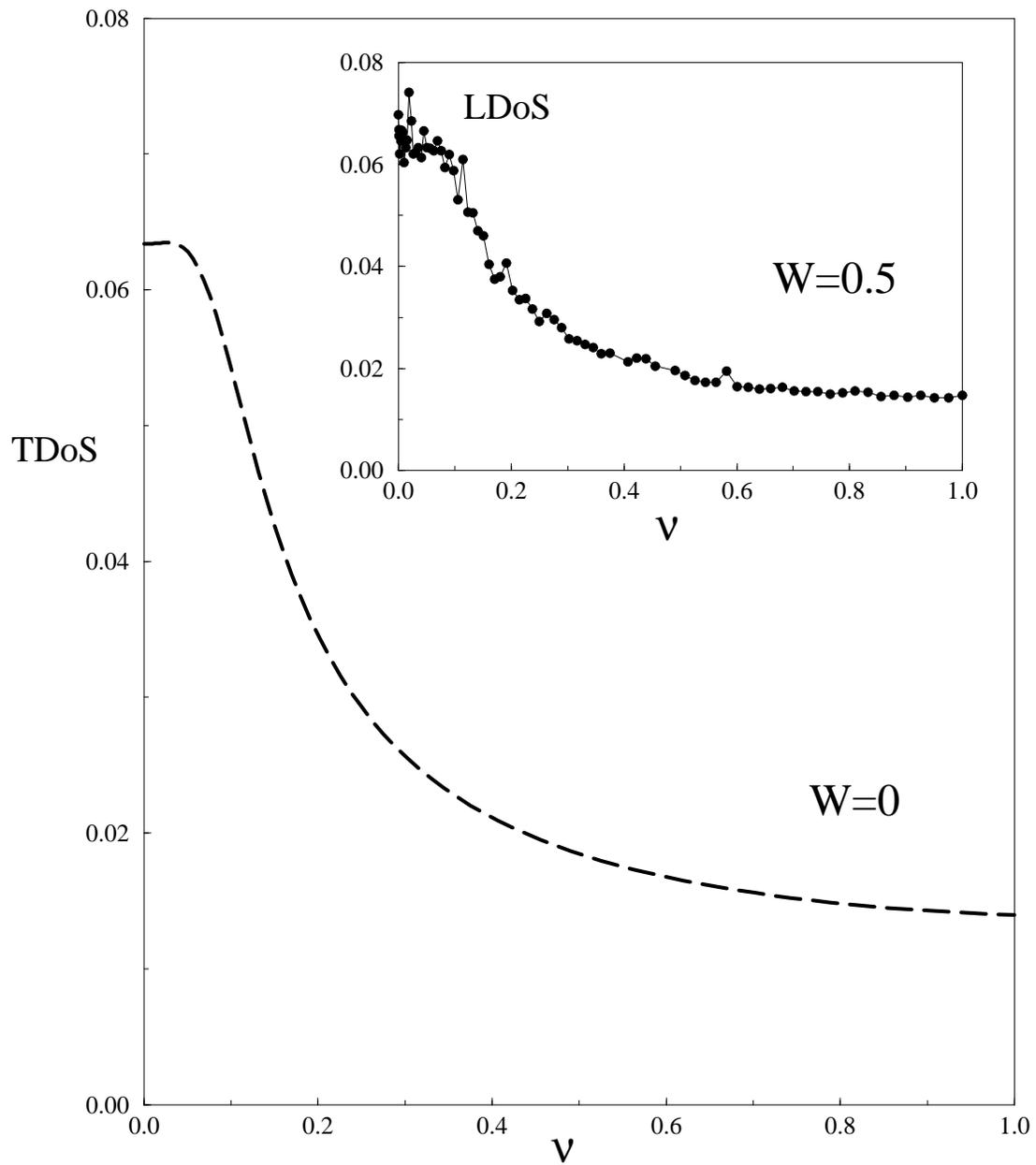
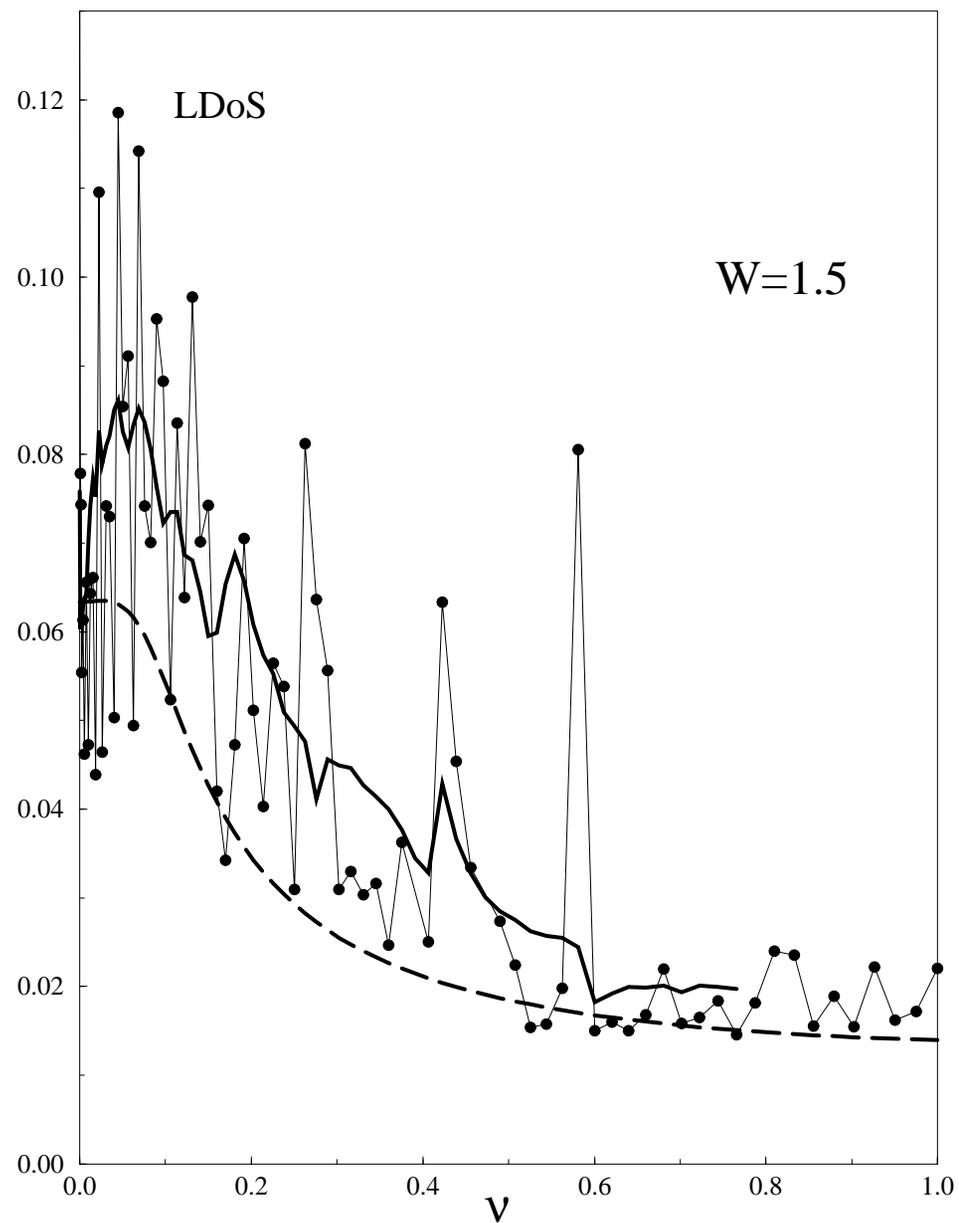
\end{document}